\documentclass[prl,twocolumn,floatfix,superscriptaddress]{revtex4-1}
\usepackage{graphicx}
\usepackage[columnwise]{lineno}
\usepackage{amsmath}
\usepackage{hyperref}
\usepackage{bibunits}
\usepackage{xcolor}

\newlength{\onecolfig}
\newlength{\twocolfig}
\newcommand{\unit}[1]{\,\mbox{#1}}
\newcommand{\mHz}{\unit{mHz}}
\newcommand{\Hz}{\unit{Hz}}
\newcommand{\kHz}{\unit{kHz}}
\newcommand{\MHz}{\unit{MHz}}

\newcommand{\mW}{\unit{mW}}

\newcommand{\nm}{\unit{nm}}

\newcommand{\uK}{\unit{$\mu$K}}
\newcommand{\s}{\unit{s}}

\newcommand{\ms}{\unit{ms}}
\newcommand{\us}{\unit{$\mu$s}}
\newcommand{\ns}{\unit{ns}}

\newcommand{\dB}{\unit{dB}}
\newcommand{\dg}{^\circ}

\newcommand{\gnd}{\ensuremath{^1\mathrm{S}_0^{m_J=0}}}
\newcommand{\exc}{\ensuremath{^3\mathrm{P}_1^{m_J=-1}}}
\newcommand{\fth}{\ensuremath{\delta_\mathrm{3D}}}
\newcommand{\ftw}{\ensuremath{\delta_\mathrm{2D}}}
\newcommand{\fvs}{\ensuremath{\delta_\mathrm{s}}}
\newcommand{\dfl}{\ensuremath{\delta_\mathrm{L}}}
\newcommand{\fl}{\ensuremath{f_\mathrm{L}}}
\newcommand{\fc}{\ensuremath{f_\mathrm{c}}}
\newcommand{\dca}{\ensuremath{\delta_\mathrm{ca}}}

\newcommand{\fa}{\ensuremath{f_\mathrm{a}}}
\newcommand{\fp}{\ensuremath{f_\mathrm{pr}}}

\newcommand{\fd}{\ensuremath{f_\mathrm{c}'}}

\newcommand{\sr}{\ensuremath{^{88}\mathrm{Sr}}}
\newcommand{\gt}{\ensuremath{g^{(2)}}}

\newcommand{\gtz}{\ensuremath{g^{(2)}(0)}}
\newcommand{\fthd}{\ensuremath{f_{\mathrm{3D}}}}

\begin{document}

\title{Continuous momentum state lasing and cavity frequency-pinning with laser-cooled strontium atoms}

\author{Vera M.~Schäfer}
\affiliation{JILA, NIST, and Department of Physics, University of Colorado, Boulder, CO, USA.}
\affiliation{Max-Planck-Institut f\"ur Kernphysik, Saupfercheckweg 1, 69117 Heidelberg, Germany}
\author{Zhijing Niu}
\affiliation{JILA, NIST, and Department of Physics, University of Colorado, Boulder, CO, USA.}
\author{Julia R.K.~Cline}
\affiliation{JILA, NIST, and Department of Physics, University of Colorado, Boulder, CO, USA.}
\author{Dylan J.~Young}
\affiliation{JILA, NIST, and Department of Physics, University of Colorado, Boulder, CO, USA.}
\author{Eric Yilun Song}
\affiliation{JILA, NIST, and Department of Physics, University of Colorado, Boulder, CO, USA.}
\author{Helmut Ritsch}
\affiliation{Institut f\"ur Theoretische Physik, Universit\"at Innsbruck, A-6020 Innsbruck, Austria}
\author{James K.~Thompson}
\email{jkt@jila.colorado.edu}
\affiliation{JILA, NIST, and Department of Physics, University of Colorado, Boulder, CO, USA.}

\date{\today}


\maketitle
{\bf\noindent
Laser-cooled gases of atoms interacting with the field of an optical cavity are a powerful tool for quantum sensing and the simulation of open and closed quantum systems. 
They can display spontaneous self-organisation phase transitions \cite{Black2003,Brennecke2007,Domokos2002,Baumann2010,leonard2017supersolid,Kroezelev2018,schuster2020supersolid,sauerwein2023engineering,helson2023density}, time crystals \cite{HemmerichTimeCrystal2021}, new lasing mechanisms \cite{Kruse2003,Norcia2016b,norcia2018frequency}, squeezed states \cite{Hosten2016,Braverman2019,Thompson18dBSqueezing} for quantum sensing \cite{Greve2022,Pedrozo-Penafiel2020,robinson2024direct},  protection of quantum coherence \cite{norcia2018cavity,davis2019photon,luo2023cavity,young2024observing}, and dynamical phase transitions \cite{muniz2020exploring,young2024observing}. 
However, all of these phenomena are explored in a discontinuous manner due to the need to stop and reload a new ensemble of atoms. Here we report the observation of hours-long continuous lasing from laser-cooled $^{88}$Sr atoms continuously loaded into a ring cavity.
The required inversion to produce lasing arises from inversion in the atomic momentum degree of freedom \cite{Courtois1994}, a mechanism related directly to self-organization phase transitions \cite{Black2003,Brennecke2007,Domokos2002,Baumann2010,leonard2017supersolid,Kroezelev2018,schuster2020supersolid,sauerwein2023engineering,helson2023density} and collective atomic recoil lasing \cite{Kruse2003,Slama2008}, both of which were previously only observed in a cyclic fashion compared to the truly continuous behavior here.
Further, the sensitivity of the lasing frequency to cavity frequency changes is 120 fold suppressed due to an atomic loss mechanism, opening an interesting new path to compensate cavity frequency noise for realizing narrow frequency references \cite{Meiser2009,Bohnet2012a,Norcia2016b,winchester_magnetically_2017}.
This work opens the way for continuous cavity QED quantum simulation experiments as well as continuous superradiant lasers.
}

\renewcommand{\refname}{Main text}

To achieve the sufficiently cold temperatures and high phase-space-densities necessary for quantum simulation, sensing and lasing, atoms are typically cooled by sequentially applying periods of different types of atomic cooling, causing the experiments to operate in a naturally pulsed fashion.
Developing continuous cold atom sources would greatly advance sensing \cite{Bothwell2022}, extend the length of quantum simulations and increase quantum gate depths that can be achieved with neutral atoms.
Recent such advances include the continuous production of Bose Einstein Condensate \cite{Chen2022}, continuous transport of atoms in a lattice \cite{Okaba2024,Cline2022}, continuous loading into a high finesse optical ring cavity \cite{Cline2022} towards continuous superradiance, and continuous replenishment of a large ytterbium tweezer array for quantum computation and simulation \cite{norcia2024iterative}.
Lasing in cold atoms has been observed by establishing optical inversion on narrow-linewidth atomic transitions \cite{Meiser2009,Bohnet2012a,Norcia2016b} with the potential to realize robust active frequency references.
Lasing has also been realized using Raman transitions between atomic internal ground states\cite{hilico1992operation,guerin2008mechanisms,Bohnet2012a,VrijsenKasevich2011,weiner2012superradiant}, inversion on a virtual ground state \cite{Gothe2019}, and Mollow gain on two-level optical transitions \cite{wu1977observation,grynberg1993central,guerin2008mechanisms,sawant2017lasing}.
The strong coupling between the atoms and light field and their inter-dependency also lead to feedback mechanisms that can stabilise otherwise unstable states \cite{bohnet2014linear,Bohnet2012Relax, Wolf2023}.\\
\begin{figure*}[!ht]
\centering
\includegraphics[width=\twocolfig]{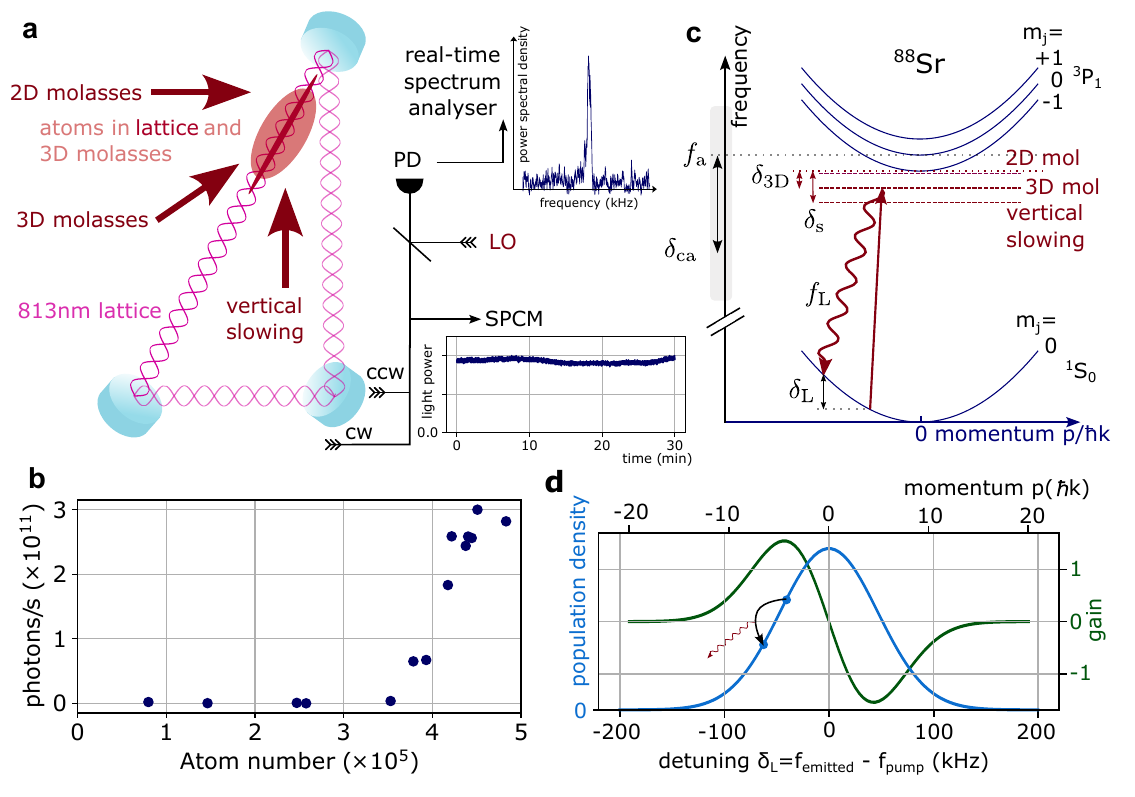}
\caption{{\bf Experimental setup:}
(a) Atoms are continuously loaded into a $813\nm$ lattice inside a high-finesse ring cavity from a 3D molasses (red cloud).
The lattice consists of two counter-propagating 813\nm\ beams (pink standing wave), and can also be used to transport the atoms along the cavity axis \cite{Cline2022}.
Several cooling lasers overlap with the atoms in the cavity mode: the 2D molasses, 3D molasses and vertical slowing (vs) beams.
Light exiting the cavity is coupled into a fibre and can be analysed either in the time domain on a Single Photon Counting Module (SPCM) or in the frequency domain on a real-time spectrum analyser via a heterodyne beatnote with a local oscillator (LO) beam derived from the same laser as the cooling beams.
Both the clockwise (cw) and counter-clockwise (ccw) direction of the cavity are monitored.
(b) The lasing shows a clear threshold behaviour with respect to the number of atoms interacting with the cavity mode.
(c) All three cooling lasers interact with the $7.5\kHz$-wide $689\nm$ \gnd\ to $^3\mathrm{P}_1$ transition (\fa, to $m_j=0$), which has an excited state Zeeman splitting of $\Delta f=\pm 1.2\MHz$.
The lasers are red-detuned from \exc\ by $\ftw=-80\kHz$, $\fth=-900\kHz$ and $\fvs=-1.6\MHz$ respectively.
The detuning of the bare cavity resonance frequency \fc\ is defined with respect to the atomic resonance frequency \fa\ as $\dca=\fc-\fa$.
Absorption of a cooling photon at frequency $f_\mathrm{3D}$ and subsequent emission of a photon at $f_\mathrm{L}$ with detuning $\dfl=f_\mathrm{L}-f_\mathrm{3D}$ into the cavity mode changes the momentum of the atoms by $\sim\hbar k$.
(d) The momentum state distribution of the atoms is defined by the Maxwell-Boltzmann distribution (MBD).
It describes the probability distribution for the frequency of the emitted photons, where the gain is proportional to the gradient of the MBD.
At the standard deviation of the MBD the momentum-state inversion, and therefore the gain, is maximal, leading to a recoil-induced-resonance (RIR, \cite{Courtois1994}).
}
\label{fig:setup}
\end{figure*} 
\begin{figure*}[!ht]
\centering
\includegraphics[width=\twocolfig]{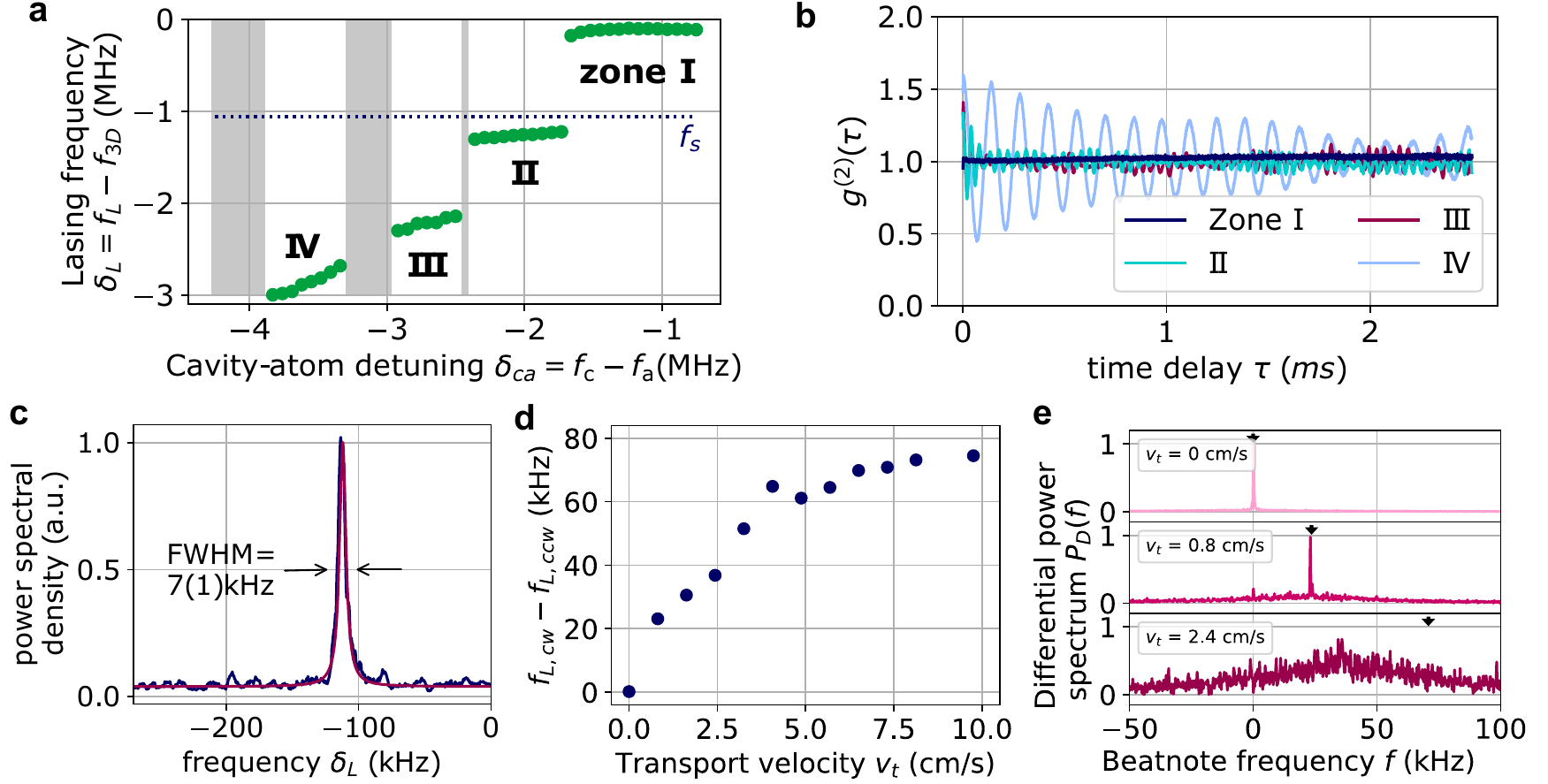}
\caption{{\bf Characterisation of the emitted light:}
(a) When scanning the bare cavity detuning \fc, four zones of light emission can be identified, differing by the frequency of the emitted light \dfl.
Between zones II\&III and zones III\&IV there are dark zones with no light emission (grey shaded areas).
The cooling lasers closest in frequency are the 3D molasses at $\dfl=0$ and the slowing beam at $f_\mathrm{s}$, indicated via the dotted line.
(b) $g^{(2)}$ correlation function for the different zones.
For zone I $g^{(2)}(\tau\to 0) = 1$ and $g^{(2)}(\tau) = 1$ throughout, confirming coherent continuous-wave laser emission.
For zones II-IV amplitude oscillations cause oscillations in $g^{(2)}(\tau)$, which become larger in amplitude and periodicity for higher-up zones.
(c) The beatnote of the lasing light in zone I with a local oscillator derived from the cooling laser has a linewidth of $\mathrm{FWHM}=7(1)\kHz$.
(d) Frequency difference of the light emitted into cw ($f_\mathrm{L,cw}$) and ccw ($f_\mathrm{L,ccw}$) directions when transporting the atoms along the cavity axis at different velocities.
The transport velocity is defined as $v_\mathrm{t} = \delta_\mathrm{t} \lambda_{813}$.
For up to $1\,\mathrm{cm/s}$ the frequency difference follows the expected Doppler shift. 
Above this velocity the measured frequency difference decreases again, as the atoms are no longer pinned inside the lattice and therefore do not travel with the lattice velocity anymore.
(e) The beatnote between the cw and ccw emitted light has a $<18(1)\Hz$ FWHM linewidth for a stationary lattice (Fourier limited by the measurement time) and $<200 \Hz$ for small transport velocities $v_\mathrm{t} \lesssim 0.8\,\mathrm{cm/s}$.
For faster transport $v_\mathrm{t}\gtrsim 2.4\,\mathrm{cm/s}$ coherence between the two lasing directions is lost.
The black arrows on top of each plot indicate Doppler shift frequencies.}
\label{fig:char}
\end{figure*}
We continuously load strontium atoms into a high finesse ring cavity \cite{Cline2022} where they are trapped using an 813~nm intracavity optical lattice, see Fig.\,\ref{fig:setup}a. 
The laser cooling required for loading the atoms is primarily accomplished using a continuous 3D red molasses at wavelength 689~nm.
There is an additional vertically oriented slowing beam at 689~nm to facilitate capture into the 3D molasses.
Once a threshold number of atoms is reached, we observe continuous light emission from the cavity, also at 689\nm\, lasting for hours, see Fig.\,\ref{fig:setup}b. 
The threshold atom number for lasing is $N\approx 300,000$.
At this atom number, the collective dispersive cavity shift $NU_0$, with single atom light shift $U_0/2\pi=\frac{(g/2\pi)^2\dca}{\dca^2+(\gamma/2\pi)^2}=12(2)\Hz$, is much larger than the cavity linewidth $\kappa=2\pi\times 50 \kHz$, leading to strong non-linear effects and self-organisation of the atoms \cite{Domokos2002,Ritsch2013}.
Here, the cavity coupling is $g=2\pi\times3.5\kHz$, the atom-cavity detuning $\dca\sim 1\MHz$ and the excited state linewidth $\gamma=2\pi\times7.5\kHz$.\\
The appearance of lasing in this system is surprising at first, as lasing requires inversion between quantum states and there is no directly apparent mechanism by which this is established.
Raman lasing is excluded since $^{88}$Sr, the isotope of strontium used here, has a single ground state, see Fig.\,\ref{fig:setup}(c). 
Lastly, Mollow gain (for example, via a 3 photon process) is excluded based on the observed frequency of the emitted light \cite{wu1977observation,grynberg1993central,guerin2008mechanisms,sawant2017lasing}. 
However, the laser cooling itself creates a thermal ensemble with more atoms at low momentum states and fewer atoms at higher momentum states. 
Thus, one expects inversion in momentum space.
This allows for a two-photon Raman gain solely between momentum states, sometimes called a recoil-induced-resonance (RIR) \cite{Courtois1994}, see Fig.\,\ref{fig:setup}(d). 
Such lasing has been observed for short durations using a Bose-Einstein Condensate \cite{Kruse2003,Slama2008}  and in the bad-cavity or superradiant limit where the cavity linewidth $\kappa$ is much larger than the gain medium's linewidth, referred to as collective atomic recoil lasing (CARL \cite{Kruse2003}). 
Here, the atoms spontaneously form an atomic density grating and laser cooling both continuously replenishes the atoms and introduces single particle repumping of the atoms back to lower momentum states to continuously maintain inversion in momentum space. 
This allows for the continuous generation of laser light.

Our system operates in a cross-over regime in which the cavity linewidth is comparable to or smaller than the gain linewidth \cite{Norcia2016b}.  
Yet, the lasing frequency does not follow the cavity frequency but remains constant over a large range of cavity frequencies.
This is achieved by a self-regulated atomic loss mechanism that keeps the dressed cavity frequency resonant with the lasing frequency.

The frequency of the emitted  light exhibits four different regimes of behavior that depend on the bare cavity frequency's detuning $\dca =\fc-\fa $ from the atomic transition frequency \fa.
We label these regimes as zones I to IV in Fig.~\ref{fig:char}a. 
Here the emitted light frequency \fl\ is expressed via its detuning $\dfl=\fl - \fthd $ from the frequency of the 3D molasses laser light \fthd . 
As \fc\ is scanned, the lasing frequency undergoes discontinuous jumps, in some cases by more than 1~MHz.

\begin{figure}[!ht]
\centering
\includegraphics[width=\onecolfig]{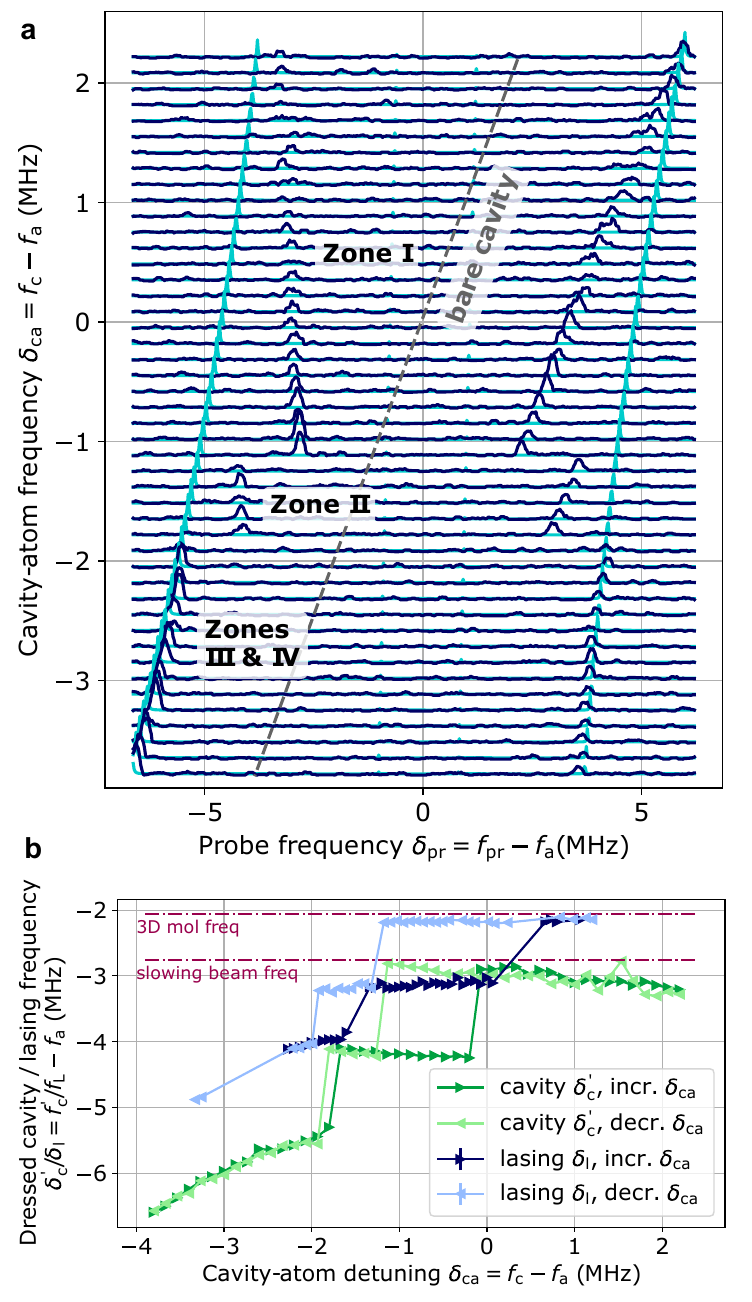}
\caption{{\bf Pinning of the cavity frequency:} 
(a) Dressed cavity frequency measured from the transmission of a weak probe beam through the cavity for different cavity-atom detunings.
The observed avoided crossing (dark blue) does not correspond to the one expected for constant atom number (simulation in turquois).
Instead, in zones I and II the dressed cavity frequency \fd\ is pinned: in zone I it stays almost constant ($\Delta\fd < 100\kHz$) over a range of $>3\MHz$ of \fc, which would normally correspond to a change of the dressed cavity frequency of $\Delta\fd = 1.3\MHz$. 
This pinning of the cavity frequency is achieved by a self-regulated reduction of the atom number.
In zones III and IV no reduction of the atom number occurs.
(b) Hysteresis and excited state population: Frequency of the dressed cavity and the emitted light for different scan directions of the bare cavity frequency \fc.
A strong hysteresis of the width and transition-frequency between zones is visible.
The light emission in a zone can be sustained for longer when adiabatically changing \fc\ and therefore only a small change in atom number per time is necessary.
The dressed cavity frequency is measured after excited state population has decayed to the ground state, leading to a frequency offset compared to its frequency during lasing.
From this offset the excited state population during lasing can be inferred.
}
\label{fig:vrs}
\end{figure}

Each zone exhibits different pulling coefficients of the light frequency with respect to changes in the bare cavity frequency $p_\mathrm{c}=d \fl/d f_\mathrm{c}$. 
Remarkably, lasing in zone~I is highly insensitive to the cavity frequency, changing by $<50\kHz$ as the cavity frequency is changed by $>3\,\MHz$, corresponding to a cavity pulling coefficient of only $p_\mathrm{c}= 8(2)\times 10^{-3}$. 
As we will show, this insensitivity to the cavity resonance frequency emerges from a lasing-induced atom loss mechanism that stabilizes the dressed cavity resonance frequency. 
 
We identify zone I as being associated primarily with the 3D molasses lasers. 
The emitted light is only $\sim100$~kHz to the red of the 3D molasses frequency.
Lasing in zone I stops immediately when the 3D molasses cooling/pumping laser is switched off, but can persist up to 700~ms after the vertical slowing beam is switched off, until the atom number has dropped below lasing threshold.
The 3D molasses pulling coefficient is $p_{\mathrm{3D,ss}}=0.9(1)$ in steady-state and $p_{\mathrm{3D,rt}}=0.31(5)$ in real-time, indicating that at short timescales the 3D molasses is in stronger competition with other mechanisms to determine the lasing frequency.
The real-time pulling coefficients were measured by applying sudden changes in a given laser frequency and observing the change in the light frequency at short time scales of $<200\us$. 
This differs from steady-state pulling coefficients since changes in laser frequencies also change the number of atoms at longer time scales. 
We identify zones III and IV as being primarily associated with the slowing beam.
For instance, lasing in zone III requires the vertical slowing beam to be on, but can persist several ms without the 3D molasses cooling light.
Our theoretical analysis of the lasing mechanism focuses on zone I, where the lasing is strongest, most robust and has the most narrow linewidth.

The measured Glauber second order correlation function $g^{(2)}(\tau)$ of the light is shown for each zone in Fig.~\ref{fig:char}b. 
In zone I, $g^{(2)}(\tau)= 1$ ($g_\mathrm{I}^{(2)}(0)= 1.01(6)$), consistent with coherent laser light emission. 
In zones II to IV we observe oscillations in $g^{(2)}(\tau)$ and $g_\mathrm{II}^{(2)}(0)=1.3(1)$, $g_\mathrm{III}^{(2)}(0)=1.4(1)$, $g_\mathrm{IV}^{(2)}(0)=1.6(1)$, indicating the light is subthermal, but exhibits superpoissonian fluctuations in its intensity with characteristic frequencies of $7-20\kHz$.  
These measurements indicate that the observed light emission is not simply due to incoherent single-particle scattering of light into the cavity mode.
 
The light's measured linewidth in zone I  continuously narrows with increasing lasing intensity. 
Close to the jump to zone II, the light reaches a $\mathrm{FWHM}=7(1)\kHz$, see Fig.\,\ref{fig:char}c.
This is considerably narrower than the cavity linewidth $\kappa$ as one expects for lasing. 
The linewidth is comparable to the $7.5\kHz$ linewidth of the $689\nm$ transition, though we do not assign any physical importance to this. 
In contrast, in zones II-IV the linewidth FWHM$\sim 100\kHz$, actually exceeds the cavity linewidth.
We also note that the 813~nm optical lattice is not necessary for achieving lasing though it helps with achieving a sufficiently high atom number to reach the lasing threshold. 
The threshold atom number can also be reached with only atoms in the 3D molasses, and lasing in zone I has been observed with both the clockwise and counter-clockwise lattice beams switched off.

We observe that light is emitted into the the ring cavity in both the clockwise (cw) and counterclockwise (ccw) directions.
The light in the two directions appears to be coherent with each other with zero frequency difference (see data at $v_\mathrm{t}=0$ in Fig.~\ref{fig:char}d and e) and a relative linewidth FWHM $18(1)$~Hz that could simply be due to relative path length noise in the heterodyne detectors.

To further study the role of the atomic gain medium in the lasing, we broke the symmetry of the coupling to the two modes by using the 813~nm lattice to continuously transport the atoms along the cavity axis at a fixed velocity $v_t$. 
For transport velocities $<1$~cm/s, the relative frequency between cw and ccw corresponds to the relative Doppler shift that one would expect for a single atom moving at $v_t$ and emitting 689~nm light into both directions at the same frequency in its reference frame. 
However, for $v_t>1$~cm/s, the relative frequency does not grow as fast as one would predict from this simple model, and the relative frequency of the two directions approaches 80~kHz, while the relative linewidth grows to ~50 kHz.

\begin{figure*}[!ht]
\centering
\includegraphics[width=\twocolfig]{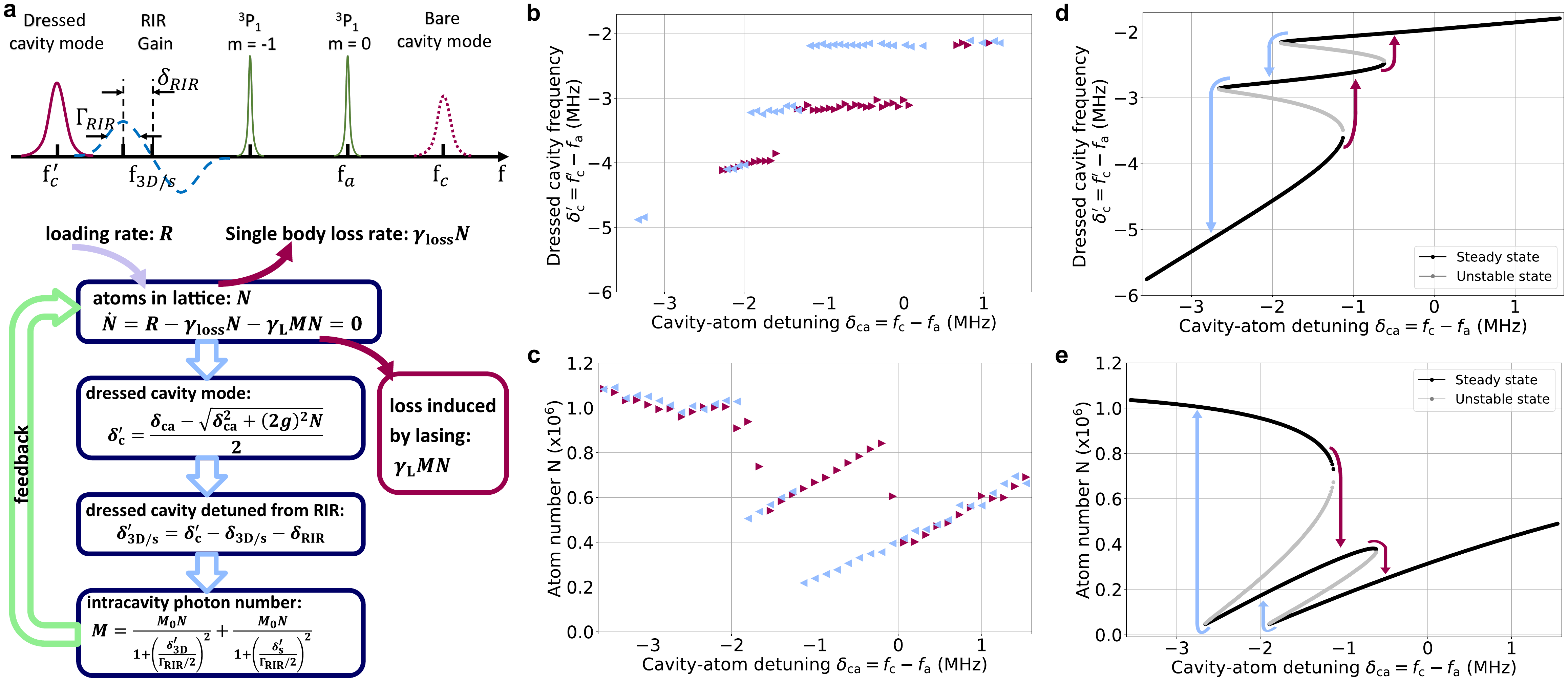}
\caption{{\bf Phenomenological model:}
(a) We can model the observed behaviour using rate equations.
The 3D molasses gives rise to a fixed loading rate R, balanced by normal atom loss from the lattice proportional to the atom number N, as well as an additional loss mechanism proportional to the emitted light with photon number M.
For a fixed cavity frequency these processes are in equilibrium, leading to a steady-state atom number N.
The atom number affects the dressed cavity frequency, which in turn affects the number of photons inside the cavity through its resonance condition.
Finally the photon number is also influenced by the width and frequency of the RIR resonance with respect to the dressed cavity frequency.
The final lasing frequency is a compromise between the dressed cavity frequency and the raw RIR peak.
This model reproduces the emergence of different zones of stable light emission with different cavity pulling coefficients and modulation of the atom number leading to pinning of the cavity frequency, as shown in the data in (b)\&(c) and corresponding simulations in (d)\&(e) (red: increasing \dca\ in time, light blue: decreasing \dca\ in time).
}
\label{fig:model}
\end{figure*} 

Having more fully characterized the nature of the observed lasing, we wish to return to the remarkable fact that the lasing in zone I persists and remains at such a constant frequency even though the bare cavity frequency is changed by $>3\MHz$.  This range is much larger than  the cavity linewidth $\kappa/(2\pi)=50\kHz$.  i.e. the lasing occurs even while its frequency is vastly off-resonant with \fc.
However, the atoms in the cavity can dress the bare cavity mode leading to an effective dressed cavity frequency \fd\ that can differ substantially from the bare cavity resonance \fc\ \cite{Thompson1992}. The lasing then occurs into the dressed cavity mode.
To investigate the dressed cavity frequency we drive the cavity with a weak probe beam of frequency \fp.
The probe beam overlaps spatially and spectrally with the emitted light, and is not visible during lasing, as the lasing intensity is much larger than the probe intensity.
Therefore we switch off all cooling lasers for $100~\mu$s, roughly three times the excited state lifetime, before applying the probe beam.
At this point the probe can be detected because all lasing has ceased, while the atom number should be unaffected, as the lattice lifetime ($\tau\approx 1\s$) is much longer.
The observed dressed cavity frequency \fd\ displays strong deviations from its expected trajectory, see Fig.\,\ref{fig:vrs}a, that can be explained only if the atom number is also varying as we change the bare cavity frequency \fc.
We indeed observe a change in atom number using fluorescence imaging of the atoms in the lattice.

In zone I over $80\%$ of the atoms are observed to be expelled from the cavity, in such a way that the dressed cavity frequency remains relatively constant over a range of $>3\MHz$ of \fc.
This drop in atom number is entirely self-regulated, and \fc\ is the only external parameter that is changed.
The atom loss could be fuelled by heating from the RIR pumping mechanism, and is discussed in more detail in the methods.
The position of the zone jumps is strongly hysteretic, see Fig.\,\ref{fig:vrs}b, and the cavity frequency has to be changed slowly to observe the full extent of a lasing zone.
A comparison between \fd\ and the lasing frequency shows identical zone jumps, but an offset between the lasing frequency and the dressed cavity frequency.
This difference is due the cooling lasers having been switched off for $100\us$ in the measurement of the dressed cavity frequency, leading to a decay of all excited state population by the time \fd\ is measured.
The characteristic interaction strength of the atoms with the cavity that leads to the cavity mode dressing is $\propto \sqrt{(N_\mathrm{g}-N_\mathrm{e})}2 g$ where $2 g$ is the single-particle vacuum Rabi frequency.
This leads to a larger vacuum Rabi splitting at the time of the dressed cavity frequency measurement relative to when the lasing is happening.
From the deviation of these two frequencies the excited state population during lasing can be inferred. 
It corresponds to roughly $30\%$ throughout.
The probe transmission can also be observed in real-time during lasing on a spectrum analyser by sweeping the probe frequency at a fixed rate.
These measurements confirm that during lasing, the lasing and dressed cavity frequency are identical to within $\sim 10\kHz$.

We constructed a phenomenological model that qualitatively captures both the observed pinning of the lasing frequency in zone I and the observed hysteretic behavior of the lasing when varying the bare cavity resonance frequency.
RIR gain favours lasing at its maximum gain frequency, corresponding to the peak momentum-state inversion. 
For instance, in zone I the measured radial atomic temperature of $10(2)\uK$ predicts that the RIR gain is maximal at a frequency $-50\kHz$ to the red of the 3D molasses frequency -- in rough, though imperfect agreement with the observed $-100\kHz$ red detuning of the lasing light.
However, in our system the predicted RIR gain linewidth is not much narrower than the cavity linewidth so the lasing can be pulled to the dressed cavity resonance frequency.  
When lasing, the atom number in the cavity $N$ depends on the lasing strength parameterized by the average number of intracavity photons $M$. 
In turn, this means that the dressed cavity frequency is tuned by the strength of the lasing.
We capture this complex dynamics using the rate equation model shown in Fig.\,\ref{fig:model}.
In this model all parameters are extracted from measurements and no free floating fit parameters were used, see the Methods section.
To capture multiple zones of lasing, we must introduce RIR gain resonances associated with both the 3D molasses and slowing lasers. 
We numerically solve for the solutions of the coupled equations shown in Fig.~\ref{fig:model}d and e. 
The black curves are the solutions stable to small perturbations and the grey curves are the solutions unstable to small perturbations.
The model qualitatively reproduces both the observed hysteresis and the frequency pinning, as well as the stepwise suppressed cavity pulling coefficients in zones II and I compared to zones III and IV.
The separate zones are due to competition between the different cooling lasers, and the number of zones increases with the introduction of more cooling lasers at different frequencies.

In summary, we have observed and explained the emergence of a narrow-linewidth continuous-wave lasing mechanism that through self-regulated feedback on the atom number has a 120 fold suppressed sensitivity to the bare cavity frequency.
We have shown how competition between different cooling lasers leads to the emergence of distinct phases of lasing with different characteristics, and how self-organisation of strongly-correlated atoms can lead to steady-state expulsion of more than $80\%$ of the atoms from the cavity.
Our lasing mechanism is coherent between both the cw and ccw modes of the ring cavity and persists even when transporting the atoms along the cavity axis at small enough speeds.

In conceptually related work with cold atoms, a four-wave mixing process requiring phase matching \cite{guerin2008mechanisms} places the atoms in a superposition of momentum states, leading to spatial self-organisation that can be identified with the phases of the dissipative Dicke model \cite{Black2003,Brennecke2007,Georges2018,guerin2008mechanisms}.
The self-organization can be viewed as a density grating that collectively enhances the scattering of light into the cavity that lasts for some short amount of time. 
In a recoil-induced resonance such as here (and the bad-cavity limit of CARL), the atoms are also placed in a quantum superposition of momentum states leading to spontaneous formation of a moving atomic density grating that can be viewed as the mechanism for scattering of the molasses pump light into the cavity. 
However, here the atoms collectively emit light into the cavity and are then reset to states of lower momentum via single particle photon scattering during the laser cooling (i.e. as is the case in standard three or four level lasers).
In contrast, typical self-organization physics experiments operate without entropy removal via single-particle scattering using essentially a collective four-wave mixing processes that can be mapped to a pseudo-spin Hamiltonian in certain limits.

By switching to the $^{87}\mathrm{Sr}$ isotope, which has a doubly forbidden $1.3\mHz$ transition well in the bad cavity regime, our system can be developed further to build a continuous-wave superradiant laser, an active \mHz-linewidth frequency reference with important metrological applications.
Our experiment also demonstrates how continuous cold atom experiments can lead to the emergence of new phenomena that involve hysteretic or bi-stable behaviour and are therefore difficult to reach in pulsed experiments.
The self-regulated pinning of the cavity frequency over a range of several \MHz\ might have interesting future applications in metrology.


\bibliographystyle{apsrev4-2}
\bibliography{mybib}{}





\section*{METHODS}

\subsection*{Loading and transport}

$^{88}\mathrm{Sr}$ atoms are continuously loaded into a lattice inside a ring cavity of finesse $\mathcal{F}=33,000$ (at 689\nm, s-polarisation), single-atom cooperativity $C=0.16$ and cavity coupling $g=3.5(2)\kHz$.
Up to $1.1\times 10^6$ atoms are loaded in steady-state, and cooled to a temperature of $10\uK$.
For loading, the atoms are heated in an oven, slowed in a Zeeman slower and then gradually cooled via a vertical chain consisting of a blue ($461\nm$) 2D MOT, a red ($689\nm$) 2D MOT, a $689\nm$ 2D molasses and finally a $689\nm$ 3D molasses overlapping spatially with the lattice \cite{Cline2022}.
In addition a vertical slowing beam (also $689\nm$) reduces the vertical velocity of the atoms.
The two MOT stages are separated from the lattice region by a baffle with a small hole for the atoms, but light from the 2D molasses, 3D molasses and vertical slowing beam spatially overlap with the atoms in the lattice.
The three cooling lasers have respective detunings of $\ftw=-80\kHz$, $\fth=-900\kHz$ and $\fvs=-1.6\MHz$ from the $\gnd$ to $\exc$ transition.
The 2D molasses beams are retro-reflected and parallel to the table, the 3D molasses beams are also retro-reflected and span all three dimensions of space, where all beams have an approximate $45\dg$ angle to the table. 
The slowing beam is almost orthogonal to the table and only propagates upwards.
The cavity axis into which the atoms are loaded has a $70\dg$ angle with respect to the table.
The cavity has a linewidth of $\kappa/2\pi=50\kHz$ at 689\nm\ and is locked to the 689\nm\ laser.
The quantisation axis is defined by a magnetic field $30^\circ$ out-of-plane relative to the cavity axis, leading to mixed polarisations for all cooling lasers.
This magnetic field leads to a $\Delta f=\pm 1.2\MHz$ Zeeman splitting of the excited states in $^3\mathrm{P}_1$.
The cavity also supports a lattice produced by two counterpropagating $813\nm$ beams, which are locked in frequency to the ring cavity.
The lattice has a depth of $150\uK$, corresponding to an axial frequency of $210\kHz$ and a radial frequency of $440\Hz$.
A travelling lattice to transport the atoms along the cavity axis is produced by introducing a frequency difference between the two lattice beams.

\subsection*{Homodyne analysis}

Both the cw and ccw output mode of the ring cavity are fibre coupled, filtered for 689\nm\ light and analysed on an SPCM.
Typical count rates on the SPCM during lasing are $1.5\MHz$.
For the \gt\ measurement, the SPCM output is recorded and timetagged on an ADC and the \gt\ correlation function is estimated via 
\begin{align}
g^{(2)}(\tau) = \frac{\sum_{i=1}^{i_\mathrm{max}-\tau} n_i n_{i+\tau}}{\left(\sum_{i=1}^{i_\mathrm{max}-\tau}n_i\right)^2}\left( i_\mathrm{max}-\tau \right)
\end{align} 
where $n_i$ are the counts detected in bin $i$, $\tau$ is the time delay from the beginning of the measurement and $i_\mathrm{max}$ is the total number of bins.
The dead-time of the SPCM is $22\ns$, and the bin time was chosen to be $300\ns$.
For coherent light $\gt(0)=1$, while for Fock states $\gt(0)=0$ and for thermal light $\gt(0)=2$. 
Since we are using a single SPCM and no beam-splitter, the dead-time of the SPCM means that coincidence counts arriving within the dead-time are not detected.
For too short bin times this artificially reduces \gtz\ below one.
Due to the cavity ringdown time of $t_\mathrm{rd}=3.2\us$ most coincidence counts are nevertheless detected in this setup without the need for a beam-splitter and two SPCMs.
For too large bin times the influence of other correlations, such as intensity fluctuations, on \gtz\ increases.
The $300\ns$ bin time was chosen as a compromise between the two effects.\\
For measuring the dressed cavity frequency, a weak probe beam is inserted into the cavity.
Since the probe beam overlaps spatially and spectrally with the lasing light, it is not visible during lasing.
To measure \fd\ during lasing all cooling lasers are switched off for $100\us$, during which the excited state decays and the lasing ceases.
After 100\us\ the probe beam is switched on and its transmission is measured for a single probe frequency \fp\ and cavity frequency \fc.
The cooling lasers are then switched on again to reload atoms and the measurement is repeated for a different \fp, leading to one horizontal trace for a fixed \fc.
Finally the cavity frequency is changed and the measurement is repeated.

\subsection*{Real-time heterodyne analysis}

For analysing the frequency of the emitted light, the fibre-coupled cavity output is overlapped on a non-polarising beam splitter with a local oscillator (LO, $\approx1\mW$ power) derived from the same laser as the cooling lasers.
The frequency of the LO is shifted such that its beatnote with the lasing light is at around $5-8\MHz$.
Both outputs of the NPBS are guided to the two ports of a differential photodiode, and the relative intensities are carefully matched using a glass pick-off to achieve maximum cancellation of common-mode noise.
The differential PD signal is then amplified by 50\,\dB, leading to a signal up to $\approx15\dB$ above noisefloor at peak lasing intensity.
The beatnote is analysed on a real-time spectrum analyser, and can be time-resolved to about $30\us$.
This setup can also be used to measure \fd\ during lasing, as the lasing light has a narrower linewidth than the cavity.
For this purpose the probe beam is swept in frequency over a range $>100\kHz$, at a rate of about $1\Hz$ during lasing.
The dressed cavity frequency is then visible as a periodically appearing broader resonance on top of the continuous signal of the lasing beatnote.
This measurement also shows the real-time shifting of the dressed-cavity frequency when switching off the cooling lasers.

\subsection*{Recoil-induced resonance}
Recoil-induced-resonances (RIR) provide a gain mechanism in cold atom systems via population inversion of the momentum states \cite{Courtois1994}.
In ring cavities, this can lead to a collective instability -- collective atomic recoil lasing (CARL) \cite{Kruse2003}.
In our system pumping is provided by the cooling lasers via RIR.
For modeling this, we look at the radial momentum of the atom cloud, as the atoms are tightly confined and in the Lamb-Dicke regime in the axial direction.
In the radial direction the confinement is weaker, and the atoms can move within little pancakes orthogonal to the cavity axis, so that they experience a recoil corresponding to the photon's momentum when absorbing/emitting a photon.
We assume a thermal distribution of the momentum states of the atoms along the radial direction, leading to the probability distribution being described by the Maxwell-Boltzmann distribution
\begin{equation}
\rho_p(p) = \frac{1}{\sqrt{2\pi mk_BT}}e^{-p^2/2mk_BT}
\end{equation}
with $m$/$p$ the mass/momentum of a \sr\ atom, Boltzmann constant $k_B$, and $T$ the radial temperature of the atom cloud.
The frequency difference for light emitted from atoms at different momentum scales quadratically, $\Delta f=p^2/2mh$.
The population inversion between two different momentum states can then be expressed as $\Delta\rho_p(p,\delta p)=\rho_p(p+\delta p)-\rho_p(p)$, where the momentum transfer $\delta p=n\hbar k$ depends on the number of photons emitted in one recoil process.
The magnitude of this inversion corresponds to the gain of the RIR mechanism.
In our setup there are three approximately retro-reflected pump lasers covering all three spatial dimensions.
We can therefore assume that in average the momentum transfer between absorbed pump photon and emitted photon in the radial directions equals one photon momentum $n=1$.
The light emitted by the atoms during RIR is shifted in frequency corresponding to the momentum gain of the atoms.
Converting the gain from momentum into frequency units yields
\begin{align}\label{RIR_equ}
&\rho_f(\Delta f) =  \sqrt{\frac{m}{ 2\pi k_BT}} \frac{\lambda}{n} \\
& \left[ e^{-(\frac{2\pi m\Delta f}{n k}+\frac{n \hbar k}{2})^2/2mk_BT}- e^{-(\frac{2\pi m\Delta f}{n k}-\frac{n \hbar k}{2})^2/2mk_BT} \right] \nonumber
\end{align}
where $k=2\pi/\lambda$ is the wave-vector of the emitted photon orthogonal to the cavity axis and $\Delta f$ is the frequency difference of the emitted light from the absorbed (cooling) light.
For an atom cloud of $T=10\uK$ the maximum gain therefore occurs at $\Delta f =50\kHz$.

\subsection*{Temperature of the atoms}
The temperature of the atoms trapped in the lattice is measured via fluorescence imaging of the thermally expanding cloud, yielding $10(1)\uK$. 
As the fluorescence of the atoms in the 3D molasses is more intense than that of atoms in the lattice, first the 3D molasses cooling lasers are switched off for $70\ms$ to allow for the atoms to gravitationally fall out of the imaged area, while only atoms trapped in the lattice remain.
$70\ms$ after the 3D molasses, the lattice beams are switched off and the atomic cloud is imaged after different expansion delays.
Since the RIR mechanism transfers atoms to a higher momentum state, the lasing-induced loss mechanism would most intuitively stem from heating of the atoms.
However no change of the temperature of the atoms is observed as the lasing intensity increases.
It would be possible that the atoms are not in a thermal equilibrium state, but that the lasing-induced heating mechanism is most efficient for already hotter atoms, that get heated out of the lattice in a run-away manner before the cloud can thermalise, because the RIR gain maximum for hotter atoms is closer to the lasing frequency.
At the same time colder atoms get cooled more efficiently due to more favourable Doppler shifts.
The hotter atoms would then no longer be confined in the lattice by the time the temperature can be measured.
While this theory qualitatively explains our observations, we do not have any experimental evidence for it, and can therefore not be certain what the nature of the lasing-induced loss mechanism is.
Density dependent mechanisms such as collisions or formation of molecules have been excluded experimentally, as no dependence of the atom loss on atom density has been observed.

\subsection*{Pulsed lasing}

\begin{figure}[!ht]
\centering
\includegraphics[width=\onecolfig]{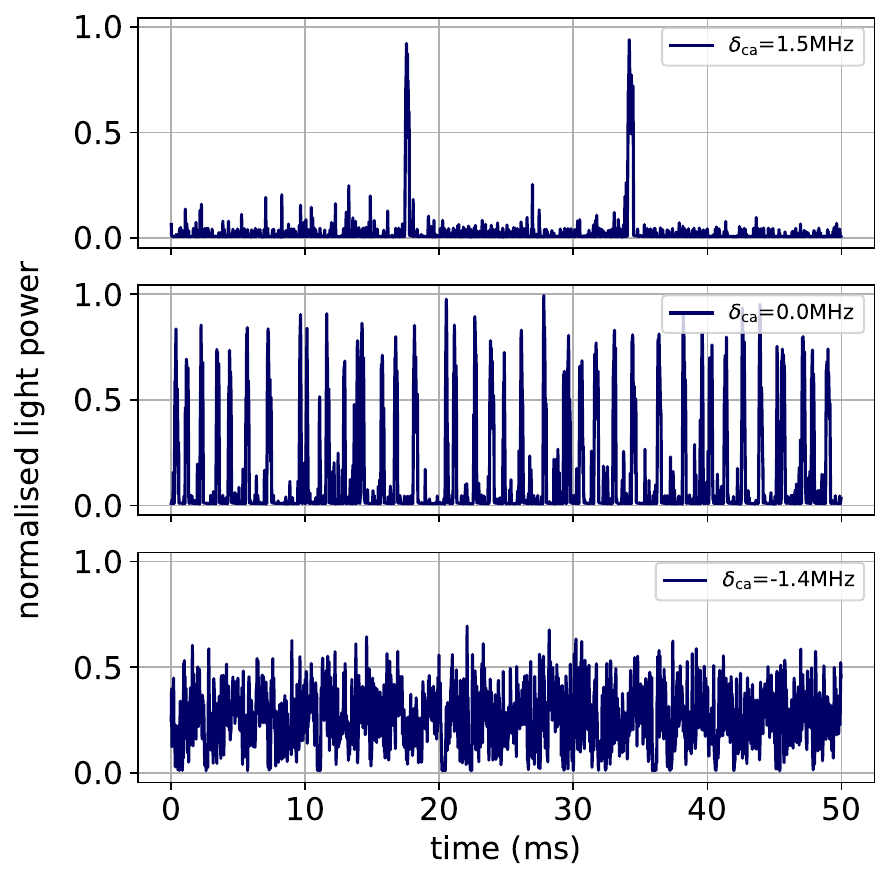}
\caption{{\bf Pulsed lasing:}
The lasing is not continuous throughout zone I, but for positive \dca\ only occasional pulses occur.
For smaller \dca, where more atoms have to be ejected from the cavity to keep the dressed cavity resonant with the lasing light the pulses are more frequent.
For even smaller \dca\ the lasing is continuous.
}
\label{fig:pulsed_traces}
\end{figure}

The lasing in zone I is not continuous throughout, but operates in a pulsed regime depending on the cavity detuning \dca, see Fig.\ref{fig:pulsed_traces}.
The measured linewidth in the pulsed regime is broadened compared to the cw lasing linewidth.
Measurements of the linewidth, pulling coefficients etc. were usually taken at $\dca\approx -1.3\MHz$, close to the transition between zones I and II, where the lasing is most intense and narrow-linewidth.
The exact position of the jump between zones depends on the reloading rate of atoms into the lattice, which depends on the number of atoms in the 3D molasses and therefore the individual laser frequencies and intensities.

\subsection*{Phenomenological model}
We consider three different atomic loading and loss mechanisms in our rate equation model:
$^{88}\mathrm{Sr}$ atoms are constantly filling up the 1D lattice at a loading rate $R = 2\times 10^7$  atoms/s, which is measured from vacuum Rabi splitting when turning on the lattice over an existing equilibrium 3D molasses.
Atoms are lost via single-body loss characterized by the rate $\gamma_\mathrm{loss} = 19$/s, which is calculated from the loading rate $R$ and the steady state atom number $N = 1.1\times 10^6$ without lasing.
The steady-state atom number $N$ is also extracted from vacuum Rabi splitting.
The second loss mechanism $\gamma_\mathrm{L} M N$ is associated with the recoil lasing. 
During lasing, the steady-state atom number will be the solution of the rate equation
\begin{equation}\label{rate_equ}
\dot{N} = R - \gamma_\mathrm{loss} N - \gamma_\mathrm{L} M N = 0
\end{equation}
The reference intracavity photon number $M_0 = 1045$ photons/atom is measured from the SPCM count rate at the edge of Zone I and Zone II where the lasing is maximal in power.
The lasing-induced-loss rate $\gamma_\mathrm{L} = 8.93 \times 10^{-6}$ $\unit{photon}^{-1} \cdot\unit{atom}^{-1}$ can be calculated from eqn.\,\ref{rate_equ}.
The number of emitted photons during lasing depends on the absorption of the pumping light, therefore the intracavity photon number M is strongly affected by the dressed cavity frequency $\fd$ and the detuning of the pumping light ($\delta_\mathrm{3D}^{'}$ and $\delta_{s}^{'}$) from the RIR gain: 
\begin{equation}
M = \frac{M_0 N}{1+(\frac{\delta_\mathrm{3D}^{'}}{\Gamma_\mathrm{RIR}/2})^2} + \frac{M_0 N}{1+(\frac{\delta_\mathrm{s}^{'}}{\Gamma_\mathrm{RIR}/2})^2}
\end{equation}
where 
\begin{equation}
\delta_\mathrm{3D/s}^{\prime} = f_\mathrm{a} + \frac{\fc - \sqrt{\fc^2 + (2g)^2 N}}{2} - (f_\mathrm{3D/s} + \delta_\mathrm{RIR})
\end{equation}

with RIR gain FWHM $\Gamma_\mathrm{RIR}=50 \kHz$ and shift $\delta_\mathrm{RIR}= 100\kHz$ calculated from eqn.\,\ref{RIR_equ}. 
The intracavity photon number $M$ in turn influences the steady-state atom number $N$, establishing a negative feedback loop.

The qualitative agreement between the equilibrium solutions of this system and our experimental observations is evident.
There are overlapping regions of stable solutions, connected by unstable solutions, which cause the observed strong hysteresis.
Notably, the cavity pulling coefficient is suppressed for zones I and II, but recovers for zones III and IV where no significant atom number modulation occurs that would impact the dressed cavity resonance.
The model also reproduces the substantial reduction of the atom number within zones I and II, a critical aspect for the maintenance of lasing and the pinning of the cavity resonance frequency.
The emergence of these distinct zones can be attributed to the competitive dynamics between different cooling lasers.
The introduction of additional cooling lasers at varying frequencies leads to a proportional increase in the number of observed and calculated zones. 














\section*{Data availability}
\noindent The data that support the plots within this paper and other findings of this study are available from the corresponding author upon reasonable request.

\end{document}